\documentclass[aps,pra,showkeys,twocolumn]{revtex4}
\usepackage{graphicx}
\usepackage{amsmath, amsthm, amssymb}
\allowdisplaybreaks[4]
\newcommand{\vect}[1] {\mathbf{#1}}
\newcommand{\dif} {\mathrm{d}}
\newcommand{\up} {\uparrow}
\newcommand{\down} {\downarrow}

\newcommand{\ket}[1] {\lvert #1\rangle}

\newcommand{\weta} {\widetilde{\eta}}

\begin{document}

\title{S-wave contact interaction problem: A simple description}
\author{Shina~Tan}
\affiliation{James Franck Institute and Department of Physics,
  University of Chicago, Chicago, Illinois 60637}

\begin{abstract}
The s-wave contact interaction problem has a very simple structure
and a simple and straightforward description. This is not a standard paper,
and technicalities are avoided. A \textit{bare} and \textit{simple}
picture of the problem is presented, in the
hope that it will be helpful to our theoretical study of the problem. 
\end{abstract}

\maketitle
The s-wave contact interaction, or s-wave resonant interaction, is
purely a quantum mechanical effect, and has no classical counterpart,
unlike the Coulomb interaction. This concept (s-wave contact interaction)
however has such
a broad range of applications that numerous theoretical and experimental
papers are published, which are closely related to it. In this regard,
it is not unlike the Coulomb interaction.

Two nonrelativistic particles in three-dimensional space
are said to have s-wave contact interaction if they
satisfy the following conditions:
\begin{itemize}
\item they have no direct interaction, if their distance \textit{or}
relative orbital angular momentum is nonzero;
\item the s-wave component (ie, spherically symmetric component)
of their \textit{relative} wave function obeys a simple
\textit{linear} boundary condition at zero distance, as below.
\end{itemize}

For simplicity, we consider particles of \textit{equal} mass $m$,
and use the natural unit $\hbar=1$. Generalizations to unequal
masses and/or restoration of $\hbar$ is easy.

We first omit the center-of-mass degree of freedom, and concentrate on
the relative motion. The Schr\"{o}dinger equation is
\begin{equation}\label{eq:Schrodinger}
i\frac{\partial}{\partial t}\psi=-\frac{1}{m}\nabla^2\psi
~~~~~~(r>0),
\end{equation}
where $\psi=\psi(\vect r, t)$, $\vect r$ is the relative wave vector
of the two particles, and $t$ is time. $r=\lvert\vect r\rvert$
is the distance between the particles.

At short distances $r\rightarrow0$, we can omit the energy, and obtain an
even simpler equation, $-\nabla^2\psi=0$. Different partial waves,
associated with different angular momentum quantum numbers $l$, are decoupled:
$\psi=\sum_{l=0}^{\infty}\psi_l$, such that
\begin{equation*}
-\frac{1}{r^2}\frac{\partial}{\partial r}
r^2\frac{\partial}{\partial r}\psi_l+\frac{l(l+1)}{r^2}\psi_l=0,
\end{equation*}
and thus $\psi_l=A_l(\hat{\vect r})r^l+B_l(\hat{\vect r})r^{-l-1}$,
where $\hat{\vect r}\equiv\vect r/r$ is the direction of $\vect r$.
\textit{Since there is no interaction in channels of nonzero angular
momenta, $B_l(\hat{\vect r})=0$ for $l>0$.} In the s-wave channel,
however, the interaction is defined by a linear \textit{constraint}
on $A_0$ and $B_0$:
\begin{equation}\label{eq:constraint}
A_0+\frac{B_0}{a}=0,
\end{equation}
and here enters the \textit{scattering length} $a$; we assume it is a nonzero
constant; it is either positive or negative, or even
infinity (the so-called \textit{unitarity limit}).
\textit{$a$ is the only parameter characterizing the s-wave contact
interaction}. Equation~\eqref{eq:constraint}, as simple as it is, is
of \textit{paramount importance} in the s-wave contact interaction problem.
Note also that \textit{the s-wave contact interaction can not
exist between identical fermions}; for other particle pairs
(including a pair of fermions at different internal states), such
interaction is allowed.

Now the wave function is generally of the form
\begin{equation}
\psi(\vect r, t)=B_0(t)\left(\frac{1}{r}-\frac{1}{a}\right)+O(r)
~~~~(r\rightarrow 0).
\end{equation}

To facilitate the study of much more complicated cases, in which
a few or many particles participate in the s-wave contact interactions,
we introduce the short-range \textit{selectors} $\lambda(\vect r)$
and $l(\vect r)$ \cite{Tan0505200} [this $l(\vect r)$ happens to involve the same English
letter as the angular momentum, unfortunately],
whose role is to selectively pick up the short-range
behavior of the wave function (or some other functions that may arise in
the problem). These selectors are possible if we
have the general notion that mathematical tools (as well as
any \textit{physical} concepts) can be freely reshaped to suit our needs:
\begin{itemize}
\item truth value (namely the proposed theory, in its range of validity,
must not contradict our experience);
\item coherence and simplicity (Ockham's razor);
\item ease to handle (namely the logical path from theory to experience
should be shortened as much as possible).
\end{itemize}

Here are the basic properties of these selectors:
\begin{subequations}
\begin{equation}
\lambda(\vect r)=l(\vect r)=0~~~~(\vect r\neq 0),
\end{equation}
\begin{eqnarray}
\int\dif^3r\lambda(\vect r)&=1,~~~\int\dif^3r\frac{\lambda(\vect r)}{4\pi r}&=0,\\
\int\dif^3r l(\vect r)&=0,~~~\int\dif^3r\frac{ l(\vect r)}{4\pi r}&=1,
\end{eqnarray}
\begin{equation}
\int\dif^3r\lambda(\vect r)\hat{\vect r}=\int\dif^3r l(\vect r)\hat{\vect r}=0.
\end{equation}
\end{subequations}
These selectors have momentum representations:
\begin{subequations}
\begin{eqnarray}
\Lambda(\vect k)&=\int\dif^3r\lambda(\vect r)\exp(-i\vect k\cdot\vect r),\\
L(\vect k)&=\int\dif^3r l(\vect r)\exp(-i\vect k\cdot\vect r),
\end{eqnarray}
\end{subequations}
and they have properties
\begin{subequations}
\begin{align}
\Lambda(\vect k)&=1~~~~(\lvert\vect k\rvert<\infty),\\
\int\frac{\dif^3k}{(2\pi)^3}\frac{\Lambda(\vect k)}{k^2}&=0,\\
\Lambda(-\vect k)&=\Lambda(\vect k);\\
L(\vect k)&=0~~~~(\lvert\vect k\rvert<\infty),\\
\int\frac{\dif^3k}{(2\pi)^3}\frac{L(\vect k)}{k^2}&=1,\\
L(-\vect k)&=L(\vect k).
\end{align}
\end{subequations}
These equations are \textit{not} contradictory, because
\begin{align*}
\int\frac{\dif^3k}{(2\pi)^3}\frac{\Lambda(\vect k)}{k^2}
&\neq\lim_{K\rightarrow\infty}\int_{\lvert\vect k\rvert<K}
\frac{\dif^3k}{(2\pi)^3}\frac{\Lambda(\vect k)}{k^2},\\
\int\frac{\dif^3k}{(2\pi)^3}\frac{L(\vect k)}{k^2}
&\neq\lim_{K\rightarrow\infty}\int_{\lvert\vect k\rvert<K}
\frac{\dif^3k}{(2\pi)^3}\frac{L(\vect k)}{k^2},
\end{align*}
in contrast with the traditional notion about integrals.

These two selectors span a linear space, called the \textit{selector space}.

Finally, we define the important $\eta$-selector, which lies in the selector space,
\begin{subequations}\begin{align}
\widetilde{\eta}(\vect r)&=\lambda(\vect r)+\frac{l(\vect r)}{4\pi a},\\
\eta(\vect k)&=\Lambda(\vect k)+\frac{L(\vect k)}{4\pi a},
\end{align}\end{subequations}
and it satisfies the important equation
\begin{equation}
\int\dif^3r\widetilde{\eta}(\vect r)\left(\frac{1}{r}-\frac{1}{a}\right)=0,
\end{equation}
but for functions that are nonsingular at $\vect r=0$,
$\widetilde{\eta}(\vect r)$ behaves like the delta function.

Equation~\eqref{eq:constraint} can now be written as
\begin{equation}\label{eq:constraint2}
\int\dif^3r\widetilde{\eta}(\vect r)\psi(\vect r, t)=0,
\end{equation}
which will play a crucial role in the s-wave contact interaction problem.

We now return to Eq.~\eqref{eq:Schrodinger}. Because the wave function
has a singular term $B_0/r$, the kinetic energy operator produces
a term $4\pi B_0\delta^{(3)}(\vect r)/m$, which has to be canceled,
if we want to make the Schr\"{o}dinger equation valid for all $\vect r$.
The simplest way to do this is to use the $L$-selector,
\begin{equation}\label{eq:zeta=0}
i\frac{\partial}{\partial t}\psi(\vect r, t)
=-\frac{\nabla^2}{m}\psi(\vect r, t)-\frac{1}{m}\delta^{(3)}(\vect r)\int
\dif^3r'l(\vect r')\psi(\vect r',t),
\end{equation}
which must be solved in conjunction with Eq.~\eqref{eq:constraint2} in order
to determine the wave function.

There is also a simple way to \textit{combine} the above equations
\begin{multline}\label{eq:zeta}
i\frac{\partial}{\partial t}\psi(\vect r, t)
=-\frac{\nabla^2}{m}\psi(\vect r, t)\\+\frac{1}{m}\delta^{(3)}(\vect r)
\int\dif^3r'\bigl[-l(\vect r')+\zeta\widetilde{\eta}(\vect r')\bigr]\psi(\vect r', t),
\end{multline}
whose solutions necessarily satisfy \textit{both} Eq.~\eqref{eq:constraint2} and
Eq.~\eqref{eq:zeta=0}. Here $\zeta$ is any \textit{nonzero} length.

The selectors $-L+\zeta\eta$, for all values of $\zeta$, form a straight line in the selector
space, and I call it \textit{Olshanii-Pricoupenko straight line}. Simple mathematics
shows that Eq.~(1) of Ref.~\cite{Olshanii2002PRL} exactly corresponds to this line.

\textit{Note that our $\Lambda$-selector is a completely different thing from
the momentum scale ``$\Lambda$'' in \cite{Olshanii2002PRL} or in
any other papers}.

The momentum scale ``$\Lambda$'' in \cite{Olshanii2002PRL}
is related to the length $\zeta$: ``$\Lambda$''$=1/a-4\pi/\zeta$.

The Olshanii-Pricoupenko pseudopotential is now expressed in terms of the selectors as
\begin{equation}
V^{\zeta}(\vect r, \vect r')=\frac{1}{m}\delta^{(3)}(\vect r)
\bigl[-l(\vect r')+\zeta\widetilde{\eta}(\vect r')\bigr],
\end{equation}
and, in particular, when $\zeta=4\pi a$, we get back to the Fermi pseudopotential,
which can now be expressed as
\begin{subequations}
\begin{equation}
V(\vect r, \vect r')=\frac{4\pi a}{m}\delta^{(3)}(\vect r)\lambda(\vect r'),
\end{equation}
whose momentum representation is simply
\begin{equation}
V(\vect k, \vect k')=\frac{4\pi a}{m}\Lambda(\vect k').
\end{equation}
\end{subequations}
It was first noted by Olshanii-Pricoupenko \cite{Olshanii2002PRL}
that the Fermi pseudopotenial is a special case of many equivalent choices.

Such equivalence is due to the fact that the wave function
satisfies the constraint Eq.~\eqref{eq:constraint2}.

The new insights that we have are:
\begin{itemize}
\item the short-range behavior of the wave function
is confined in a linear space (and is subject to an additional linear constraint);
\item
we can construct its \textit{dual linear space}, called the \textit{selector space},
whose elements are called the selectors;
\item
each selector corresponds to a well-defined generalized function, and transformation
between different representations (eg, between coordinate space and momentum space)
is easy and straightforward;
\item this allows us to formulate the \textit{whole} problem in a
\textit{simple}, \textit{unified}, and \textit{flexible}
way, and significantly facilitates our study of the whole system; 
\item the old forms of the pseudopotentials in terms of partial
derivatives are both inelegant and not flexible (for example, no one was ever
able to transform them to the other representations), and are no longer needed, once
the selectors are clearly defined.
\end{itemize}

We now proceed to a system of fermions in two ``opposite'' ``spin'' states,
subject to some external potential $V_\text{ext}$;
the s-wave contact interaction is present between the opposite spin states. The system
is described by two equations:
\begin{subequations}
\begin{equation}\label{eq:constraint3}
\int\dif^3r~\weta(\vect r)\psi_{\up}(\vect R+\vect r/2)
\psi_{\down}(\vect R-\vect r/2)\ket{\phi}=0,
\end{equation}
\begin{multline}\label{eq:H3}
H=\sum_\sigma\int\dif^3r\psi_\sigma^\dagger(\vect r)[-\nabla^2/2m+V_\text{ext}(\vect r)]
\psi_\sigma^{}(\vect r)\\
-\frac{1}{m}\int\dif^3r\dif^3r'\psi_{\up}^\dagger(\vect r)\psi_{\down}^\dagger(\vect r)
\psi_{\down}^{}(\vect r-\vect r'/2)\psi_{\up}^{}(\vect r+\vect r'/2)l(\vect r'),
\end{multline}
\end{subequations}
where $\ket{\phi}$ is any physically allowed state. We can of course combine the
two equations as previously.

It is shown in \cite{Tan0505200} that the energy of the system can be exactly expressed
in terms of the one-particle reduced density matrix only:
\begin{equation}
E=\sum_{\vect k\sigma}\eta(\vect k)\frac{k^2}{2m}n_{\vect k\sigma}
+\sum_{\sigma}\int\dif^3rV_\text{ext}(\vect r)\rho_{\sigma}(\vect r),
\end{equation}
where $n_{\vect k\sigma}$ and $\rho_\sigma(\vect r)$ are the momentum distribution
and number density distribution, respectively. Note that $\eta(\vect k)=+1$ for any finite
$\vect k$, but the peculiar behavior of $\eta(\vect k)$ upon integration allows the existence
of \textit{negative} energy states, including the two-body bound states.

Equation~\eqref{eq:constraint3} directly leads to many identities that must be satisfied
by various reduced density matrices; see \cite{Tan0505200} for details.

The same approach can also be used to study the Bose gas with s-wave contact interaction.
However, the same level of completeness will \textit{not} be reached
until we also treat the Efimov effect, \textit{inherent} in this system, clearly. This issue
was not noticed in \cite{Olshanii2002PRL}.

Generalization of this formalism to \textit{two} dimensions is easy. The relative wave
function is of the form $B_0\ln(r/a)$ plus higher order terms at short distances. We
can therefore define two selectors $\weta(\vect r)$ and $l(\vect r)$ such that
\begin{eqnarray*}
\int\dif^2r\weta(\vect r)&=1,~~~~\int\dif^2r\weta(\vect r)(2\pi)^{-1}\ln(a/r)&=0,\\
\int\dif^2r l(\vect r)&=0,~~~~\int\dif^2r l(\vect r)(2\pi)^{-1}\ln(a/r)&=1,
\end{eqnarray*}
and $\lambda(\vect r)=l(\vect r)=0$ for $\vect r\neq0$. Their Fourier transforms
can be shown to satisfy:
\begin{subequations}
\begin{equation}
\eta(\vect k)=1~~~~(\lvert\vect k\rvert<\infty),
\end{equation}
\hspace{0mm}
\begin{equation}
\int_{\lvert\vect k\rvert>K}\frac{\dif^2k}{(2\pi)^2}\frac{\eta(\vect k)}{k^2}
=\frac{1}{2\pi}\ln\frac{2e^{-\gamma}}{Ka}\doteq\frac{1}{2\pi}\ln\frac{1.1229}{Ka},
\end{equation}
\begin{equation}
L(\vect k)=0~~~~(\lvert\vect k\rvert<\infty),
\end{equation}
\begin{equation}
\int\frac{\dif^2k}{(2\pi)^2}\frac{L(\vect k)}{k^2}=1,
\end{equation}
\end{subequations}
where $\gamma=0.5772\cdots$ is Euler's constant, and $K$ is any positive momentum scale.

Now the two-component Fermi gas in two dimensions with s-wave contact interaction
satisfies two equations which look \textit{formally identical} with
Eqs.~\eqref{eq:constraint3} and \eqref{eq:H3}, except that the
three-dimensional spatial integrals are replaced by two-dimensional ones.

Using the logic similar to that in \cite{Tan0505200}, we can show that
the two-dimensional system satisfies the energy theorem
\begin{equation}
E=\sum_{\vect k\sigma}\eta(\vect k)\frac{k^2}{2m}n_{\vect k\sigma}
+\sum_{\sigma}\int\dif^2rV_\text{ext}(\vect r)\rho_{\sigma}(\vect r),
\end{equation}
which looks almost identical with its three-dimensional counterpart,
except that now the meaning of $\eta(\vect k)$ is different.

This formalism is much more straightforward than the one used in
\cite{Pricoupenko2004PRA}. Also the energy theorem is unknown to
people before.

As soon as we modify the traditional notion about integrals, in the way demonstrated
in \cite{Tan0505200} and here, the ultraviolet divergence problems
suddenly disappear.

While this approach is certainly not the first one which removes ultraviolet
divergence problems, a lot of advantages exist in this formalism. Let us compare it with
two other approaches:
\begin{itemize}
\item the traditional pseudopotential approach, which has been further
developed by Olshanii and others \cite{Olshanii2002PRL};
\item the bare coupling constants plus dimensional regularization plus renormalization
approach.
\end{itemize}
The first approach never gained the flexibility to reach a simple momentum representation.
The second approach, confined by the traditional notion of integrals, involves
complicated intermediate steps (in which the dimensionality of the system has to be
changed), divergent bare coupling constants, and subtraction procedures after the
regularizations, and is thus a long detour to a simple problem.

The s-wave contact interaction problems, in both three and
two dimensions, are now formulated easily with the short range selectors, and other
known approaches can hardly compete with it in this area. The author however does not
exclude the possibility of the appearance of a still simpler
\textit{new} approach.

Whether the ideas here can be further extended to other important systems, in particular
relativistic quantum field theories (in which the divergent bare coupling constants
plus dimensional regularization plus renormalization approach currently dominates),
remains an open question for the moment. We will get an answer to this question,
either positive or negative.

Finally, let us discuss a little bit about some ideas underlying the present
approach. Since the prospects of this approach beyond the current two systems
are still not clear, any discussion here should be regarded as rather tentative.

The traditional notion about integrals can be illustrated with a simple example,
$\int\dif^3 k/k^2$, which contains divergent contribution from large $\vect k$.
Here $\vect k$ is either momentum, or \textit{spatial vector}, or even
some other quantity. Let us assume it is momentum from this point on.
This apparent divergence seems to stem from
an assumption, namely that \textit{any short-distance physics,
including Planck scale physics, and maybe even beyond Planck momentum,
is \textbf{really} there, \textbf{actively} underlying that cloud of ultracold atoms,
even when we have not fed any particles with very large
energies}. An even deeper assumption underneath this one, is \textit{reductionism}:
\textit{short-range physics completely determines long-range physics, and the former is more
fundamental than the latter.} Whether these assumptions are valid, is of course
a question beyond the scope of this tiny article, and we shall not
attempt to discuss it further here.

The intuition that we have gained from our new generalized functions is:
short-range physics \textit{emerges} when we accelerate particles to its associated
energy scale; whether it is really there, \textit{before} we accelerate particles
to that energy scale, seems to be a philosophical question - neither provable nor
disprovable. However, in the present approach, the short-range physics appears to
be somehow \textit{activated} or maybe, more strangely, \textit{created},
after the particles are raised to high energies. Is this picture
an illusion caused by this mathematical method, or does it suggest some element of
truth?

The author is indebted to Z.~Q.~Ma and C.~Chin for some discussions.

\bibliography{contact}
\end{document}